\documentclass[11pt]{amsart}
\usepackage{amsmath}
\usepackage{amssymb}
\usepackage{amsthm}
\usepackage{latexsym}
\usepackage{tikz}
\usetikzlibrary{matrix,arrows}
\usepackage{eucal,graphicx}
\usepackage[all, knot]{xy}
\usepackage{enumerate}
\usepackage{appendix}
\usepackage{color}

\newtheorem{theorem}{Theorem}[section]

\newtheorem{corollary}[theorem]{Corollary}

\newtheorem{definition}[theorem]{Definition}
\newtheorem{example}[theorem]{Example}

\newtheorem{lemma}[theorem]{Lemma}

\newtheorem{prop}[theorem]{Proposition}
\newtheorem{remark}[theorem]{Remark}

\newtheorem{note}[theorem]{Note}
\newtheorem{fact}[theorem]{Fact}

\numberwithin{equation}{section}

\def\C{\mathbb{C}}
\def\S{\mathbb{S}}
\def\R{\mathbb{R}}
\def\E{\mathcal{E}}
\def\Rb{\textbf{$d$-RBord}}
\def\CA{\mathcal{A}}

\def\sn{\operatorname{Sym^n}}
\def\Sm{\operatorname{Sym}^*}
\def\sk{\operatorname{Sym^k}}

\def\smt{\operatorname{Sym}^2}
\def\exp{\operatorname{Exp}}
\def\Helm{\operatorname{Helm}}

\def\Cn{C^{\infty}}
\def\Cd{\mathcal{C}_{d}}
\def\Rb{$d$\text{-RBord}}
\def\H{\textbf{Hilb}}
\def\des{\operatorname{\Delta_{\Sigma}}}
\def\de{\Delta}
\def\DS{{D_{\Sigma}}}

\begin{document}

\title{Functorial Quantum Field Theory in the Riemannian setting}

\author{Santosh Kandel} 
\date{}

\maketitle

\begin{abstract}
We construct examples of functorial quantum field theories in the Riemannian setting by quantizing free massive bosons. 
\end{abstract}

\tableofcontents

\section{Introduction}

\hskip-5mm There are a number of mathematical approaches to rigorously define Quantum Field Theory. As suggested by the path integral quantization, in the framework pioneered by Atiyah, Kontsevich, Segal and many others, roughly a $d$-dimensional Riemannian QFT is a rule $E$ that assigns to a $d$-dimensional closed oriented Riemannian manifold $\Sigma$ a number $E(\Sigma)$ which depends only on the isomorphism class of $\Sigma$. More generally, E assigns to a $d$-dimensional compact oriented Riemannian manifold $\Sigma$ with $\partial \Sigma=\overline{Y_0}\sqcup{Y_1}$ Hilbert spaces $E(Y_0)$ and $E(Y_1)$ together with a bounded operator $E(\Sigma):E(Y_0)\to E(Y_1)$. The main requirement is that $E$ is local. Here by local we mean if $\Sigma$ is obtained by gluing $\Sigma_1$ and $\Sigma_2$ along the common boundary $Y$, then the contributions from $\Sigma_1$ and $\Sigma_2$ are sufficient to compute $E(\Sigma)$. 
\\[1mm]
More precisely, let $\Rb$ be the $d$-dimensional Riemannian bordism ``category'' whose objects are oriented closed $d-1$ dimensional Riemannian manifolds. We use $Y$s to denote objects of this category. A morphism $\Sigma:Y_0\to Y_1$ in $\Rb$ is a $d$-dimensional compact oriented Riemannian manifold with orientation preserving isometry $\overline{Y_0}\sqcup Y_1\to \partial\Sigma.$ Here $\overline{Y_0}$ means the intrinsic orientation on $Y_0$ is reversed. We require a morphism $\Sigma$ in $\Rb$ to have \textit{product metric} near the boundary so that we have a well defined composition of morphisms by gluing. We remark $\Rb$ is just a semicategory not a category as it does not have identity morphisms.
\\[1mm]    
Let \textbf{Hilb} be the category whose objects are separable Hilbert spaces and morphisms are continuous linear operators. We note $\H$ is a symmetric monoidal category with the monoidal structure given by the Hilbert space tensor product. 
\\[1mm]
Following Atiyah\cite{Atiyah} and Segal \cite{Seg4}, the discussion above can be summarized as
                \begin{definition}\label{Def1.1}
A $(d,d-1)$  Riemannian Functorial QFT is a functor $E: \Rb\to\textbf{Hilb}$ that maps disjoint unions into tensor products. 
                \end{definition} 
  
\hskip-5mm We will also consider a more general bordism category (\ref{MFT}), namely our objects and morphisms come with a fixed vector bundle with a metric and a compatible connection. Our main goal here is to construct a class of examples of $(d,d-1)$ Riemannian FQFTs. Naively, one can try to do path integral quantization to construct such examples. Path integral quantization heuristically means construction ``measures'' on the space of fields given by a classical field theory. 
\\[1mm]
A classical field theory on a $d$-dimensional compact oriented Riemannian manifold $\Sigma$ consists of two pieces of data:  space of fields $\mathfrak{F}(\Sigma)$ and an action functional $S$ which is a real valued function on $\mathfrak{F}(\Sigma).$ Furthermore, one requires $S$ to be local which means $S$ depends only on fields and their derivatives.
 \\[1mm]             
The space of fields $\mathfrak{F}(\Sigma)$ is space of sections of a fiber bundle $P$ over $\Sigma.$ An example of space of fields is $C^{\infty}(\Sigma)$, the space of real valued smooth functions on $\Sigma$ and an example of action functional is the energy functional \begin{eqnarray}\label{af}S(\phi)=\frac12\int_{\Sigma}d\phi\wedge\ast d\phi+m^2\ast\phi^2
\end{eqnarray}  where $\ast$ is the Hodge star operator associated to the Riemannian metric and $m$ is a positive real number. 
\\[1mm]                  
The idea behind path integral quantization is to make sense of the integrals
                         \[E(\Sigma)=\int_{\mathfrak{F}(\Sigma)}e^{-S(\phi)}\,D\phi\]
where $D\phi$ is a `` volume measure" on $\mathfrak{F}(\Sigma)$ and it is very difficult to achieve in general. Fortunately, using ideas from constructive quantum field theory \cite{GJ}, it is possible to construct ``measures'' in some linear $\sigma$-models. However, these measures, which are required to define path integrals, live on the space of distributions not on the space of fields. This does not create serious problems in some nice cases. In \cite{P}, Douglas Pickrell follows this point of view to construct a class of examples of $2$-dimensional FQFTs called $P(\phi)_2$-interaction theories. We use ideas from Pickrell\cite{P} and Segal \cite{Seg} to show

      \begin{theorem}\label{thI.1} There is a projective representation of $\Rb$ in \textbf{Hilb}.
      \end{theorem}
      
       \hskip-4.5mm Our main result is to promote this projective representation to get a FQFT when $d$ is even. More precisely, we prove  
       \begin{theorem}\label{thI.2}There exists $d$-dimensional Riemannian FQFT when $d$ is even.
       \end{theorem} 
       
       \hskip-4.5mm In fact, theorem \ref{thI.2} can be used to construct FQFT in all dimensions as follows. A closed oriented Riemannian 1-manifold $S$ induces a functor $\Rb\to (d+1)\text{-RBord}$ by taking product with $S$. When $d$ is odd, composing this functor with the one given by theorem \ref{thI.2} gives a FQFT. This shows that
       \begin{corollary} There are FQFTs for $d$ odd as well.
       \end{corollary}

\paragraph{Acknowledgements}
This paper is based on a part of the author's PhD thesis at the University of Notre Dame. The author is grateful to his advisor Stephan Stolz for the discussions, guidance and support. The author also thanks Brian Hall and Liviu Nicolaescu for their inspiring ideas and comments. The author benefited from conversation with Owen Gwilliam, Douglas Pickrell and Augusto Stoffel. It is the author's pleasure to thank Sayonita Ghosh Hajra for feedback and Max Planck Institute for Mathematics for the hospitality during the author's stay where part of this work was written.

\section{Background}\label{section2}
\subsection{Dirichlet-to-Neumann operator}    
  
Let $\Sigma$ be a $d$-dimensional compact oriented Riemannian manifold, $C^{\infty}(\Sigma)$ the space of real valued smooth functions on $\Sigma$, $\Delta_{\Sigma}$ the nonnegative Laplacian on $\Sigma$ and $m>0$.
\\[2mm]
Assume $Y=\partial\Sigma\ne\varnothing$. Let $\nu$ denote the outward unit normal vector to $Y$ and $i:Y\to \Sigma$ be the inclusion. We recall a classical fact that the Dirichlet problem  \[(\Delta_{\Sigma}+m^2)\phi=0\hskip0.05in \text{with}\hskip0.05in \phi|_{Y}=\eta\] has a unique solution $\phi_{\eta}$ in $\Cn(\Sigma)$ for all $\eta\in \Cn(Y)$ (\cite{MS}, page 193) 
            We call the unique solution $\phi_{\eta}$ to the Dirichlet problem the $Helmholtz$ $extension$ of $\eta$.

 \begin{definition}\label{Def2.1} The operator on $Y$ defined by 
  \[D_{\Sigma}\eta=\frac{\partial \phi_{\eta}}{\partial\nu}\] is called the Dirichlet-to-Neumann operator associated to the Helmholtz operator $\des+m^2.$ 
 \end{definition}

\hskip-4.5mm It is well known that $D_{\Sigma}$ is a positive elliptic pseudo differential operator of order one  as an operator on $Y$ (\cite{Tay}, chapter 7, section 11) and it has the same principal as the operator $(\Delta_{Y}+m^2)^{\frac12}$ (\cite{Lee2}, Theorem 2.1). In particular, $D_{\Sigma}$ is symmetric.
\\[1mm]
Let $\Helm(\Sigma)$ denote the space of Helmholtz solutions on $\Sigma$ and $\Phi(Y)$ denote $\Cn(Y)\oplus\Cn(Y).$ There is a map $\Helm(\Sigma) \to \Phi(Y)$ given by $\Psi\mapsto (\Psi|_{Y}, \frac{\partial\Psi}{\partial\nu}|_{Y}).$ We also note that $\Phi(Y)$ can be given a canonical symplectic structure $\omega_{\Phi(Y)}$ which is the hyperbolic form induced by the $L^2$ inner product on $\Cn(Y).$    
       
\begin{remark}The image of $\Helm(\Sigma)$ under the map above can be identified with the graph of $D_{\Sigma}:\Cn(Y)\to \Cn(Y)$ inside $\Phi(Y)$ and hence is a Lagrangian subspace of $\Phi(Y).$
 \end{remark}

\subsubsection{Gluing Dirichlet-to-Neumann operators}\label{Subsec2.1.1} 
     
Let $\Sigma_{1}$ be a $d$ dimensional oriented compact Riemannian manifold with $\partial\Sigma_{1}=Y_2\sqcup\overline{Y_1}$ where $\overline{Y_1}$ means $Y_1$ with the induced orientation which is opposite of the intrinsic orientation on $Y_1$. We call $Y_1$ is an incoming boundary and $Y_2$ an outgoing boundary.  Let $\Sigma_2$ be another $d$ dimensional oriented compact Riemannian manifold with $\partial\Sigma_{2}=Y_3\sqcup\overline{Y_2}$ and $\Helm(\Sigma_2)\circ \Helm(\Sigma_1)$ be the fiber product $\Helm(\Sigma_2)\times_{\Phi(Y_2)}\Helm(\Sigma_1).$ More precisely $\Helm(\Sigma_2)\circ \Helm(\Sigma_1)$ contains pairs $(\phi_2,\phi_1)$ where $\phi_2(\phi_1)$ is a Helmholtz solution on $\Sigma_2(\Sigma_1)$ such that their values and normal derivatives agree on $Y_2.$ Let $\Sigma_2\circ \Sigma_1$ denote the Riemannian manifold glued along $Y_2.$ It is easy to see that there is a canonical isomorphism between $\Helm(\Sigma_2)\circ \Helm(\Sigma_2)$ and $\Helm(\Sigma_2\circ \Sigma_1).$ We can rephrase this discussion in terms of Dirichlet-to-Neumann operators.  
\\[1mm]
We can write
         \[D_{\Sigma_1}= \left[\begin{smallmatrix} A & B\\ B^t &D\end{smallmatrix}\right]:\Cn(Y_2)\oplus \Cn(Y_1)\to \Cn(Y_2)\oplus \Cn (Y_1)\] where 
              \begin{align*}&A:\Cn(Y_2)\to \Cn(Y_2), B:\Cn(Y_1)\to \Cn(Y_2),\\
                                        & D:\Cn(Y_1)\to \Cn(Y_1)\hskip0.05in\text{and}\hskip0.05in B^{t}:\Cn(Y_2)\to \Cn(Y_1).
                   \end{align*}
Now, let us describe the operators $A$, $B$, $B^t$ and $D$. Given $\phi_2$ in $\Cn(Y_2)$, calculate the Helmholtz solution on $\Sigma_1$ with boundary value $\phi_2$ on $Y_2$ and vanishing boundary value on $Y_2$; then $A(\phi_2)$ is the outward normal derivative of the Helmholtz solution along $Y_2$. Given $\phi_1$ in $\Cn(Y_1)$, calculate the Helmholtz solution on $\Sigma_1$ with boundary value $\phi_1$ on $Y_1$ and vanishing boundary value on $Y_1$; then $D(\phi_1)$ is the inward normal derivative of the Helmholtz solution along $Y_1.$ Similarly we define $B$ and $B^{t}.$ The notation $B^{t}$ is used to remind us $D_{\Sigma_1}$ is a symmetric operator.
\\[1mm]         
We write \[D_{\Sigma_2}= \left[\begin{smallmatrix} K & L\\ L^t &M\end{smallmatrix}\right]:\Cn(Y_3)\oplus \Cn(Y_2)\to \Cn(Y_3)\oplus \Cn(Y_2). \]
 
\begin{lemma}\label{L2.3}\[D_{\Sigma_2\circ\Sigma_1}=\left[\begin{smallmatrix} K-L(A+M)^{-1}L^t & -L(A+M)^{-1}B\\ -B^t(A+M)^{-1}L^t &D-B^t(A+M)^{-1}B\end{smallmatrix}\right]\] as an operator from $\Cn(\partial(\Sigma_2\circ\Sigma_1))$ to itself.
\begin{proof} We will only sketch the proof. For simplicity, we assume $Y_1=\varnothing.$ We need to show \[D_{\Sigma_2\circ\Sigma_1}\psi =K-L(D_{\Sigma_1}+M)^{-1}L^t\psi\] for all $\psi\in \Cn(Y_3)$. Let $\Psi$ be the Helmholtz solution on $\Sigma_2\circ\Sigma_1$ such that $\Psi|_{Y_2}=\psi$ and let $\Psi_2$ be the Helmholtz solution on $\Sigma_2$ such that $\Psi_2|_{Y_3}=\psi$ and $\Psi|_{Y_2}=0.$ Then, $\Psi-\Psi_2$ is the Helmholtz solution on $\Sigma_2$ such that $\Psi-\Psi_2|_{Y_3}=0$ and $\Psi-\Psi_2|_{Y_2}=\phi$ where $\phi=\Psi|_{Y_2}.$ This implies
\begin{eqnarray*}
\begin{array}{lll} B\phi=D_{\Sigma_2\circ\Sigma_1}\psi-K\psi \hskip2mm\text{and}\hskip2mm B^{t}\psi=-(D_{\Sigma_1}+M)\phi.
\end{array}
\end{eqnarray*}
Now, the fact $(D_{\Sigma_1}+M)$ is invertible (Theorem 2.1, \cite{Ca}) proves the lemma.
 \end{proof} 
 \end{lemma}

\hskip-4.5mmLet $Y$ be a closed oriented Riemannian manifold, $m>0$ and $s\in\R$. The bilinear form on $C^{\infty}(Y)$ given by 
           \[<f,g>=\int_{Y}f(\Delta_{Y}+m^2)^{s}g\, dvol(Y)\] defines an inner product on $C^{\infty}(Y)$. We define the $Sobolev$ $space$ $W^s(Y)$ as the completion of $C^{\infty}(Y)$ with respect to this inner product.

             \begin{lemma}\label{Lem2.3} Let $\Sigma$ be a compact oriented Riemannian manifold and $\partial\Sigma=Y.$ Then \[\alpha_{\Sigma}=\DS(\de_{Y}+m^2)^{-\frac12}\] defines a unique continuous positive operator on $W^{\frac12}(Y).$
             
             \begin{proof} Recall that $\DS$ and $(\de_{Y}+m^2)$ both are positive pseudo differential operators of order one and both same principal symbol. This means  
              \[\DS(\de_{Y}+m^2)^{-\frac12}=I+R\] 
              where $R$ is a pseudo differential operator of order at most $-1.$ This shows  $(\de_{Y}+m^2)^{-\frac12}\DS$ is a pseudo differential operator of order zero and hence it has a unique extension to a continuous linear operator on $W^{\frac12}(Y).$ Positivity of $\alpha_{\Sigma}$ is obvious.  
             
              \end{proof}
               \end{lemma}  
                     
\hskip-4.5mm Let $\Sigma_1$ and $\Sigma_2$ be as above. Recall that $\alpha_{\Sigma_1}$, $\alpha_{\Sigma_2}$ and $\alpha_{\Sigma_2\circ \Sigma_1}$ are continuous positive operators on $W^{\frac12}(Y_3)\oplus W^{\frac12}(Y_2)$, $W^{\frac12}(Y_2)\oplus W^{\frac12}(Y_1)$  and $W^{\frac12}(Y_3)\oplus W^{\frac12}(Y_1)$ respectively.
              We want to show $\alpha_{\Sigma_1}$, $\alpha_{\Sigma_2}$ and $\alpha_{\Sigma_2\circ \Sigma_1}$ are related in the similar fashion as $D_{\Sigma_1}$, $D_{\Sigma_2}$ and $D_{\Sigma_2\circ\Sigma_1}.$
          The following definition is useful to describe this relation.

              \begin{definition}\label{Def2.5} Let $H_1$, $H_2$ and $H_3$ be Hilbert spaces,  \[\CA_1:H_2\oplus H_1\to H_2\oplus H_1\hskip0.03in \text{and}\hskip0.03in \CA_2:H_3\oplus H_2\to H_3\oplus H_2\] be positive continuous operators with continuous inverses. We write 
              \[\mathcal{A}_1=\left[\begin{smallmatrix} A & B\\ B^t &D\end{smallmatrix}\right] \hskip0.03in\text{and}\hskip0.03in  
              \mathcal{A}_2=\left[\begin{smallmatrix} K & L\\ L^t &M\end{smallmatrix}\right].\] We define 
                 \begin{equation}\label{Eq 1.1}\mathcal{A}_2\circ\mathcal{A}_1=\left[\begin{smallmatrix} K-L(A+M)^{-1}L^t & -L(A+M)^{-1}B\\ -B^t(A+M)^{-1}L^t &D-B^t(A+M)^{-1}B\end{smallmatrix}\right].
                 \end{equation}
                 \end{definition}
                 
\hskip-4.5mm As a corollary of lemma \ref{Lem2.3}, we have 
                 
                 \begin{corollary}\label{C2.4}$\alpha_{\Sigma_2}\circ \alpha_{\Sigma_1}=\alpha_{\Sigma_2\circ \Sigma_1}.$
                  \end{corollary}

\hskip-4.5mm Next, we describe the idea behind the definition \ref{Def2.5}. We will consider the finite dimensional situation. Informally, we can use $\CA_1$ and $\CA_2$  to construct Lagrangian subspaces in some symplectic vector spaces and the definition of $\CA_2\circ \CA_1$ is just a consequence of ``composing" Lagrangian subspaces. We refer to lemma \ref{L2.6} for a precise statement.

               \begin{definition} Let $V$ and $W$ be two vector spaces. A linear relation $P:V\to W$ is a subspace of $W\oplus V.$
                \end{definition} 
              \hskip-4.5mm Let $P:V\to W$ and $Q:W\to U$ be two linear relations. The composition $Q\circ P$ of linear relations is the fiber product $Q\times_{W}P$ which is defined by \[Q\circ P =\{(u,v)|\exists w\in W \hskip0.03in\text{with} \hskip0.03in (u,w)\in U\oplus W, (w,v)\in W\oplus V\}.\]

                 \begin{definition} 
                 
                 Let $(V,\Omega_{V})$ be a finite dimensional symplectic vector space and $V_{\C}$ be the complexification of $V$ and abusing the notation, $\Omega_{V}$ be the complex bilinear extension of $\Omega_{V}.$ A subspace $L$ of $V_{\C}$ is said to be a positive Lagrangian if $L$ is a Lagrangian subspace and the Hermitian form defined by 
                 \[<v,w>=-i\Omega_{V}(\overline{v},w)\] is positive definite on $L.$
                 
                 \end{definition}
                 
                \hskip-4.5mm Let $(V,\Omega_{V})$ be as above. We use the notation $\overline{V}$ to denote the symplectic vector space $(V,-\Omega_{V}).$

              \begin{lemma}\label{L2.9}
              
                Let $U$, $V$, $W$ be finite dimensional symplectic vector spaces. Let $P$ be a positive Lagrangian subspace of $V_{\C}\oplus\overline{U_{\C}}$ and $Q$ be a positive Lagrangian subspace of $W_{\C}\oplus\overline{V_{\C}}.$ Then $Q\circ P$ is a positive Lagrangian subspace of $W_{\C}\oplus\overline{U_{\C}}.$
              
              \begin{proof} Using the well known fact the composition of Lagrangian linear relations is again a Lagrangian linear relation (\cite{Ne1}, Theorem 8.3) we see $Q\circ P$ is a Lagrangian linear relation. To complete the proof we have to show $Q\circ P$ is positive Lagrangian subspace which is a simple computation. Let $(w,u)\in Q\circ P$, then there is $v\in V_{\C}$ such that $(w,v)\in Q$ and $(v,u)\in P.$ Now,
             \[i\Omega_{W_{\C}\oplus \overline{U_{\C}}}\left((\overline{w},\overline{u}),(w,u)\right)=i\Omega_{W_{\C}\oplus \overline{V_{\C}}}\left((\overline{w},\overline{v}),(w,v)\right)+i\Omega_{V_{\C}\oplus \overline{U_{\C}}}\left((\overline{v},\overline{u}),(v,u)\right)\] implies $-i\Omega_{W_{\C}\oplus \overline{U_{\C}}}\left((\overline{w},\overline{u}),(w,u)\right)$ is positive definite.
              
               \end{proof}
              \end{lemma}

                  \hskip-4.5mmLet $H$ be a finite dimensional real Hilbert space and define $V=V^{+}\oplus V^{-}$ where $V^{+}=V^{-}=H.$ Then $(V,\Omega_{V})$ is a symplectic vector space with the symplectic form the hyperbolic form defined by \[\Omega((v_1^+,v_1^-),(v_2^+,v_2^-))=<v_1^+,v_2^->-<v_2^+,v_1^->.\] Let $V_{\C}$ be the complexification of $V.$ Let $\CA:H\to H$ be a positive operator. Proof of the following lemma is obvious.

           \begin{lemma}
           
           The graph of $i\CA$ is a positive Lagrangian subspace of $V_{\C}.$
         
           \end{lemma}

           \hskip-4.5mm More generally, let $H_1$, $H_2$ and $H_3$ be finite dimensional real Hilbert spaces. Let $\CA_1: H_2\oplus H_1\to H_2\oplus H_1$ and $\CA_2: H_3\oplus H_2\to H_3\oplus H_2$ be positive operators. We write  
           \[\mathcal{A}_1=\left[\begin{smallmatrix} A & B\\ B^t &D\end{smallmatrix}\right] \hskip0.03in\text{and}\hskip0.03in
            \mathcal{A}_2=\left[\begin{smallmatrix} K & L\\ L^t &M\end{smallmatrix}\right].\] Define \[\widetilde{\CA_1}=\left[\begin{smallmatrix}A&B\\-B^t&-D\end{smallmatrix}\right]\hskip0.03in \text{and} \hskip0.03in\widetilde{\CA_2}=\left[\begin{smallmatrix}K&L\\-L^t&-M\end{smallmatrix}\right].\]

     \hskip-4.5mm It is easy to check  

    \begin{lemma}\label{L2.6} 
   
      \begin{enumerate}[(i)]
\item The subspace $L_1$ which is the graph of $i\widetilde{\mathcal{A}_1}$ is a positive Lagrangian subspace of \[(H_2\oplus H_2)_{\C}\oplus \overline{((H_1\oplus H_1)_{\C}}.\] 
\item The subspace $L_2$ which is the graph of $i\widetilde{\mathcal{A}_2}$ is a positive Lagrangian subspace of \[(H_3\oplus H_3)_{\C}\oplus \overline{(H_2\oplus H_2)_{\C}}.\]
\item  $L_2\circ L_1$ is the graph of $i\widetilde{\mathcal{A}_2\circ\mathcal{A}_1}$ where \[\CA_2\circ \CA_1=\left[\begin{smallmatrix} K-L(A+M)^{-1}L^t & -L(A+M)^{-1}B\\ -B^t(A+M)^{-1}L^t &D-B^t(A+M)^{-1}B\end{smallmatrix}\right].\] 
     \end{enumerate}
     
     \end{lemma}

\subsection{Gaussian measures and Bosonic Fock spaces}\label{2.5}
              
 Here we recall few facts about Gaussian measures on a real nuclear vector space $V$ which we use in this paper. Our main reference for Gaussian measure related facts is \cite{V}. A well known fundamental fact is that there is a canonical bijection between the set of nondegenerate centered Gaussian measures on $V^{\lor}$ and the set of inner products on $V$. Given a nondegenerate centered Gaussian measure $\mu$ on $V^{\lor}$ the corresponding inner product on $V$ is given by \[<v,w>=\int_{V^{\lor}}(v,F)\cdot(w,F)\,d\mu(F).\]  where $(v,F)$ and $(w,F)$ are the natural pairings between elements of $V$ and $V^{\lor}$ respectively. Conversely, given an inner product on $V$ we can use Milnos theorem to construct the corresponding Gaussian measure. 
 \\[2mm]                           
We follow the slogan ``almost all'' features of a nondegenerate Gaussian measure $\mu$ on $V^{\lor}$ are controlled by a subspace $H(\mu)$ of $V^{\lor}$ where $H(\mu)$ carries a Hilbert space structure. This Hilbert space is called the $Cameron$-$Martin$ space of the pair $(V^{\lor},\mu)$ (See \cite{V}, page 44 for the precise definition of Cameron-Martin space). We can think of the Cameron-Martin space as the dual of the Hilbert space which is the completion of the $V$ with respect to the inner product that defines the given Gaussian measure. 
                            
                               \begin{example} \label{Ex2.12}
                             
                                Let $Y$ be an object in $\Rb$ which means $Y$ a closed oriented $d-1$ Riemanninan manifold. Consider the   
                                inner product on $\Cn(Y)$ given by \[<f,g>_{W^{\frac12}(Y)}=\int_{Y}f(\de_{Y}+m^2)^{\frac12}g\,dvol(Y)\] 
                                Let $\mu_{Y}$ be the corresponding Gaussian measure on $D'(Y)$ the space of distributions on Y. Then the Sobolev space $W^{\frac12}(Y)$ 
                                 is the Cameron-Martin space of $\mu_{Y}.$
                               
                               \end{example} 
                                                                                  
\hskip-4.5mm One key feature of the Cameron-Martin space is that it can be used to detect whether two Gaussian measures are mutually absolutely continuous or not. Let $(V^{\lor},\mu)$ be as before and let $H(\mu)$ be its Cameron-Martin space. Let $\nu$ be another Gaussian measure on $V^{\lor}$ with the Cameron-Martin space $H(\nu)$, then $\mu$ and $\nu$ are mutually absolutely continuous if and only if $H(\mu)=H(\nu)$ as vector spaces and there is a symmetric Hilbert-Schmidt operator $S:H(\mu)\to H(\mu)$ such that \[<u,v>_{H(\nu)}=<u,v>_{H(\mu)}+<Su,v>_{H(\mu)}\] for all $u,v\in H(\mu).$ See chapter 9 in \cite{V} for details.
\\[1mm]
Another feature which interests us is that it completely characterizes the square integrable functions in the sense we explain below. 
 \\[1mm]         
Let $V$ be a nuclear space, $\mu$ a Gaussian measure on $V^{\lor}$ and $H(\mu)$ the Cameron-Martin space of $\mu$. Then there is an isomorphism of Hilbert spaces \[S:L^2(V^{\lor},\mu)\to \Sm H(\mu)^{\lor}\] which is known as \textit{Segal-Ito} isomorphism (\cite{AG}, Chapter 3). Here $\Sm H(\mu)^{\lor}$ is the Bosonic Fock space (See \ref{4.1}) of $H(\mu)^{\lor}$. If $V$ is a finite dimensional real Hilbert space then $S$ is essentially the forward heat operator. Hence we may think of $S$ as the infinite dimensional version of the forward heat operator.

          \begin{remark}\label{Re2.13}The upshot of the previous paragraph is that we do not really need Gaussian measure to describe the space of square integrable functions. It is enough to have the Cameron-Martin space at hand. 
           \end{remark}

\hskip-4.5mm We use remark \ref{Re2.13} frequently explicitly or implicitly to turn measure theoretic constructions into constructions in Bosonic Fock spaces.

\section{Construction of free theories}  
\subsection{Heuristic computation of path integrals and gluing formula for the zeta regularized determinants} We describe a heuristic path integral quantization of a free massive scalar field theory to motivate the construction of the FQFTs in this paper. We recall that we want to make sense of the integrals of the form 
\begin{eqnarray}\label{3.1}E(\Sigma)=\int_{\mathfrak{F}(\Sigma)}e^{-S(\phi)}\,D\phi
\end{eqnarray} where $S(\phi)$ is given by (\ref{af}). 
\\[2mm]
\underline{\textbf{$\Sigma$ is closed:}} First, we assume $\Sigma$ is closed. Then, using integration by parts, we can write
 \begin{eqnarray*} S(\phi)=\frac12\int_{\Sigma}\phi(\Delta_{\Sigma}+m^2)\phi\, dvol(\Sigma)
 \end{eqnarray*}
This shows $E(\Sigma)$ is a Gaussian. If $\mathfrak{F}(\Sigma)$ were a finite dimensional vector space, then $E(\Sigma)$ would be simply $\det(\Delta_{\Sigma}+m^2)^{-\frac12}.$ Hence, we  need a generalization of determinant for the infinite dimensional case and zeta regularized determinant \cite{LF} is one such generalization. This suggests $E(\Sigma)=\det_{\zeta}(\Delta_{\Sigma}+m^2)^{-\frac12}.$
\\[2mm]
\underline{\textbf{$\Sigma$ has nonempty boundary:}} Assume that $\partial\Sigma=Y.$ Let $\eta\in \Cn(Y)$ and we want to heuristically make sense of integral
\begin{eqnarray}\label{3.2}E(\Sigma)(\eta)=\int_{\{\phi\in \mathfrak{F}(\Sigma): \phi|_{Y}=\eta\}}e^{-S(\phi)}\,D\phi
\end{eqnarray}
First, we need the following lemma
\begin{lemma}\label{L3.1} If $\psi$ is a Helmholtz solution, then 
\[\int_{\Sigma}d\phi\wedge \ast d\psi+m^2\ast\phi\psi=\int_{Y}\phi D_{\Sigma}\psi|_{Y}\,dvol(Y).\]
\begin{proof} Using $d(\phi\ast d\psi)=d\phi\wedge \ast d\psi+\phi d\ast d\psi$, we have 
\begin{align*}&\int_{\Sigma}d\phi\wedge \ast d\psi+m^2\ast\phi\psi\\
                     &=\int_{Y}\left(\phi\ast d\psi\right)|_{Y}+\int_{\Sigma}\phi(-d\ast d\psi+m^2\ast\psi)\\
                     &=\int_{Y}\left(\phi\ast d\psi\right)|_{Y}
\end{align*} 
Now, \[\int_{Y}\left(\phi\ast d\psi\right)|_{Y}=\int_{Y}\phi D_{\Sigma}\psi|_{Y}\,dvol(Y)\] follows from the proof of Proposition 4.1.54 of \cite{LN}.
\end{proof}
\end{lemma}

\hskip-4.5mmNow, we go back to heuristic computation of $E(\Sigma)(\eta).$ Let $\phi\in \mathfrak{F}(\Sigma)$ such that $\phi|_{Y}=\eta.$ Let us write $\phi=\phi_{\eta}+\hat{\phi}$, where $\phi_{\eta}$ is the Helmholtz solution with boundary value $\eta$ and $\hat{\phi}$ vanishes on the boundary. Using lemma \ref{L3.1} and integration by parts, we see  
\begin{eqnarray*} S(\phi)=S(\phi_{\eta})+S(\hat{\phi})=\frac12\int_{Y}\eta D_{\Sigma}\eta\,dvol(Y)+\frac12\int_{\Sigma}\hat{\phi}(\Delta_{\Sigma}+m^2)\hat{\phi}\, dvol(\Sigma)
\end{eqnarray*}     
Hence, heuristically, $E(\Sigma)(\eta)=e^{-S(\phi_{\eta})}\det_\zeta (\Delta_{\Sigma, D}+m^2)^{-\frac12}$ where $\Delta_{\Sigma, D}$ is the Laplacian with the Dirichlet boundary condition.                          
\\[1mm]                                          
\underline{\textbf{Gluing:}} Now assume that $\Sigma=\Sigma_2\circ\Sigma_1$ in $\Rb$ where $\Sigma_1:\varnothing\to Y$ and $\Sigma_2:Y\to \varnothing.$ In particular, $\Sigma$ is closed. Then
\begin{eqnarray*}
\begin{array}{lll}&E(\Sigma_2)\circ E(\Sigma_1)\\
                           &=\det_{\zeta}(\Delta_{\Sigma_1, D}+m^2)^{-\frac12}\det_{\zeta}(\Delta_{\Sigma_2, D}+m^2)^{-\frac12}.\\
                           &\int_{\eta\in\C(Y)}e^{-\frac12 \left<(D_{\Sigma_2}+D_{\Sigma_1})\eta,\eta\right>}D\eta\\ 
                           &=\det_{\zeta}(\Delta_{\Sigma_1, D}+m^2)^{-\frac12}\det_{\zeta}(\Delta_{\Sigma_2, D}+m^2)^{-\frac12}\det_{\zeta}(D_{\Sigma_2}+D_{\Sigma_1})^{-\frac12}
 \end{array}                                                                             
\end{eqnarray*}                                                                      
                                                                                    
\hskip-4.5mm In order to have a FQFT, we want $E(\Sigma)=E(\Sigma_2)\circ E(\Sigma_1)$. Hence, we expect to have a gluing formula for the zeta regularized determinants. In other words, if we compute path integrals heuristically and expect a FQFT, then it predicts the gluing formula for the zeta regularized determinants.
\\[1mm]
Fortunately, there is a gluing formula for the zeta regularized determinants (\cite{B}, Theorem A or \cite{Lee2}, Corollary 1.3) when $d$ is even. Hence, heuristic computation of path integrals leads to $E(\Sigma)=E(\Sigma_2)\circ E(\Sigma_1).$
 \\[1mm]
In summary, heuristic computations of path integrals together with the gluing formula for the zeta regularized determinants suggests existence of functorial quantum field theories. In the rest of the paper, we show this is indeed the case. In particular, when $\partial\Sigma=Y$, we want to give a precise mathematical meaning to $E(\Sigma)$ from above. We want to think of $E(\Sigma)$ as a vector in a Hilbert space associated to $Y$ and we will see, in our approach, that we must modify $E(\Sigma)$ from above. 
                                                                                                                
\subsection{A strategy for construction of free theories} In \cite{Seg4}, Segal suggests free theories can be constructed following $\Rb\to T_{pol} \to \H.$ We heuristically follow this path. Here $T_{pol}$ is the ``category'' of ``polarized'' symplectic vector spaces whose objects are polarized symplectic vector space $(V, J)$ and a morphism $J:(V_0, J_0)\to (V_1, J)$ is a polarization of $V_1\oplus\overline{V_0}$ which is ``close'' to $J_1-J_0.$ For $Y\in \Rb$ the datum $(\Phi(Y), \sqrt{\Delta_{Y}+m^2})$ can thought as a ``polarized'' symplectic vector space. To an object $Y$ the ``functor'' $\Rb\to T_{pol}$ assigns $(\Phi(Y), \sqrt{\Delta_{Y}+m^2}).$ For a morphism $\Sigma\in\Rb$ the operator $D_{\Sigma}$ can be interpreted as a morphism in $T_{pol}.$ The other ``functor'' $T_{pol} \to \H$ can be thought as passing to the quantum theory.
    
\subsection{Construction of Hilbert space of states} 
Let $Y$ be an object in $\Rb.$ There are at least two equivalent ways to construct the Hilbert space associated to $Y.$ 
\\[1mm]
We recall one can assign the symplectic vector space $(\Phi(Y), \omega_{\Phi(Y)})$ where $\Phi(Y) =\Cn(Y)\oplus\Cn(Y)$ and $\omega_{\Phi(Y)}$ is the hyperbolic form induced by the $L^2$ inner product on $\Cn(Y).$ The underlying vector space of $\Phi(Y)$ can be thought as the Cauchy data for the classical solutions associated to the action functional (\ref{af}) on a very thin cylinder around $Y.$ We need an extra data namely a polarization to perform geometric quantization to construct a Hilbert space. We may think of graph of $i\sqrt{\Delta_{Y}+m^2}$ as a positive polarization (see section 9.5 in \cite{PS}) on $\Phi(Y).$ Now, one can assign to $Y$ the Hilbert space which is obtained by performing geometric quantization.
\\[1mm]
Since we are interested in explicit computations, we use the ideas from constructive quantum field theory (\cite{GJ}, \cite{P}) to construct the Hilbert space of states. 
         
                     \hskip-4.5mmTo an object $Y$ of $\Rb,$ we assign the Hilbert space of states  \[E(Y)=\Sm W^{\frac12}(Y)^{\lor}.\] 
            Recall from section \ref{2.5} that $E(Y)$ is the $L^2$ space of the Gaussian measure on $D'(Y)$ whose Cameron-Martin space is $W^{\frac12}(Y).$ 
            
             \hskip-4.5mm It is clear from the construction $E(\phi)=\R$, $E(\overline{Y})=E(Y)^{\lor}$, and $E(Y_1\sqcup Y_2)=E(Y_1)\otimes E(Y_2)$ where $\otimes$ is the Hilbert space tensor product.
            \begin{remark} Following \cite{P}, E(Y) can be defined as the the space of half densities of the measure class the Gaussian measure $\mu$ on $D'(Y)$ with the Cameron-Martin space $W^{\frac12}(Y).$
            \end{remark}
         
         \begin{remark} The complexification of $E(Y)$ is essentially an irreducible unitary representation of the``standard'' Heisengberg group of $\Phi(Y)$ as explained in section 9.5 in \cite{PS} or section 5.4 in \cite{Gef}.
       \end{remark}

      \subsection{Construction of Operators} 
                               
                                        Let $\Sigma:\varnothing\to Y$ be a morphism in $\Rb.$  We want to construct a vector $E(\Sigma)\in E(Y).$ Let us first explain measure theoretic idea behind the construction. We note there is a Gaussian measure $\mu_{Y}$ on the space of distributions $D'(Y)$ on Y whose Cameron-Martin space is the Sobolev space $W^{\frac12}(Y).$ Recall that $\DS$ is a positive operator on $Y$ and hence defines an inner product on $\Cn(Y)$ given by 
                               \[<f,g>_{1}=\int_{Y}f\DS g\, dvol(Y).\] Let $\nu_{Y}$ be the Gaussian measure on $D'(Y)$ corresponding to the inner product $<,>_{1}.$
                               
                                  \hskip-4.5mmBy our assumption $\Sigma$ has product metric near $Y.$ It turns out that $\DS-(\de_{Y}+m^2)^{\frac12}$ is a smoothing operator (Theorem 2.1, \cite{Lee2}). As a consequence of this fact we observe the measures $\mu_{Y}$ and $\nu_{Y}$ are mutually absolutely continuous. Now the idea is to define
                                the vector $E(\Sigma)$ is a suitable multiple of the squareroot of the Radon-Nikodym derivative $\nu_{Y}$ with respect to $\mu_{Y}.$ We will give this construction in the Fock space setting.

                              \hskip-4.5mmWe recall $\alpha_{\Sigma}=\DS(\de_{Y}+m^2)^{-\frac12}$ and  $\alpha_{\Sigma}$ is a continuous positive operator on $W^{\frac12}(Y).$ Moreover,
                          
                             \begin{lemma} $\alpha_{\Sigma}-I$ is trace class operator on $W^{\frac12}(Y).$

                            \begin{proof} Theorem 2.1 in \cite{Lee2} implies  $\DS-(\de_{Y}+m^2)^{\frac12}$ is a smoothing operator on $Y.$ This means  \[\alpha_{\Sigma}=I+R\] where $R$ is a smoothing operator. Since $R$ is smoothing the unique extension of $R$ on $W^{\frac12}(Y)$ is trace class. 
                            \end{proof}
                            \end{lemma}

       \hskip-4.5mmLet $C(\alpha_{\Sigma})=(I-\alpha_{\Sigma})(I+\alpha_{\Sigma})^{-1}$ be the Cayley Transform of $\alpha_{\Sigma}$ (see definition \ref{def10}). We observe $C(\alpha_{\Sigma})$ is Hilbert-Schmidt symmetric operator on $W^{\frac12}(Y)$ and $\|C(\alpha_{\Sigma})\|<1.$ We recall from \ref{4.1}  \[\E(C(\alpha_{\Sigma}))=\exp\left(\frac12C(\alpha_{\Sigma})\right)\in \Sm W^{\frac12}(Y)^{\lor}.\]

                   \begin{remark} The vector $\dfrac{\E(C(\alpha_{\Sigma}))}{\|\E(C(\alpha_{\Sigma}))\|}$ is  the squareroot of the Radon-Nikodym derivative of $\nu_{Y}$ with respect to $\mu_{Y}$ under the $Segal$-$Ito$ isomorphism (\cite{SK}, Theorem 3.2.24). 
                   \end{remark} 
        
\hskip-4.5mmMore generally if $\Sigma_1:Y_1\to Y_2$  is a morphism in $\Rb$, then we interpret $\Sigma$ as a morphism from $\phi$ to $Y_2\sqcup \overline{Y_1}.$ Then \[\E(C(\alpha_{\Sigma_1}))\in E(Y_1)^{\lor}\otimes E(Y_2).\]  We further identify $\E(C(\alpha_{\Sigma_1}))$ with a Hilbert-Schmidt operator $E(Y_1) \to E(Y_2)$ which will be again denoted by $\E(C(\alpha_{\Sigma_1})).$

                   \begin{prop}\label{P4.1.6}
         
                   Let $H_1$, $H_2$ and $H_2$ are real separable Hilbert spaces. Let $\CA_1$  and $\CA_2$  be continuous positive operators respectively on $H_2\oplus H_1$ and $H_3\oplus H_2$ such that ${\CA}_1-I$ and ${\CA}_2-I$ 
                  are trace class operators. Then
                       
                       \begin{equation}\label{E4.2}\mathcal{E}(C(\mathcal{A}_2))\circ \mathcal{E}(C(\mathcal{A}_1))=c(\mathcal{A}_2,\mathcal{A}_1)\mathcal{E}(C(\mathcal{A}_2\circ\mathcal{A}_1))
                       \end{equation} where  
                      
                      \begin{align*} &c(\mathcal{A}_2,\mathcal{A}_1)=\dfrac{\det(\mathcal{A}_1)^{\frac14}\cdot \det(\mathcal{A}_2)^{\frac14}}{\det(\frac{A+M}{2})^{\frac12}
                      \det(\mathcal{A}_2\circ\mathcal{A}_1)^{\frac14}}\dfrac{||\mathcal{E}(C(\mathcal{A}_1))||\cdot||\mathcal{E}(C(\mathcal{A}_2))||}{||\mathcal{E}(C(\mathcal{A}_2\circ\mathcal{A}_1))||}
                     \end{align*}
                      and the norms are taken in the Fock spaces. 
                     Here \[\CA_1=\left[\begin{smallmatrix} A & B\\ B^t &D\end{smallmatrix}\right],\hskip0.05in\CA_2=                                \left[\begin{smallmatrix} K & L\\ L^t &M\end{smallmatrix}\right]
 \hskip0.05in\text{and}\hskip0.05in
\CA_2\circ\CA_1=\left[\begin{smallmatrix} K-L(A+M)^{-1}L^t & -L(A+M)^{-1}B\\ -B^t(A+M)^{-1}L^t &D-B^t(A+M)^{-1}B\end{smallmatrix}\right]\]
               
                      \begin{proof} First of all we note that $\CA_1-I$ and $\CA_2-I$ trace class  implies $\det(\frac{A+M}{2})$ and $\det(\CA_2\circ \CA_1)$ exist.
                        Also, we note $\E$, $C$ and the composition map $\circ$ are continuous. Hence, it is sufficient to show the equation (\ref{E4.2}) holds for $\CA_1$ and $\CA_2$ such that $\CA_1-I$ and $\CA_2-I$ are finite dimensional operators. This case is done in the proposition \ref{P4.4}.
                     
                       \end{proof}
                       \end{prop}

                      \begin{definition} A projective representation $T$ of a category $\mathcal{C}$ in \textbf{Hilb} assigns to an object $C$ in $\mathcal{C}$ a Hilbert Space $T(C)$, and to a  morphism $P:C\to D$ a continuous linear operator $T(P)$ such that for any pair of morphisms $P:C\to D$ and $Q:D\to  E$ we have
                         \[T(Q\circ P)=\lambda(Q,P)T(Q)\circ T(P)\] where $\lambda(Q,P)$ is a nonzero complex number. 
                         \end{definition}

               \begin{theorem}\label{Ch5:T4.1} The assignment $Y\to E(Y)$ where $Y$ is an object in $\Rb$ and $\Sigma \to \mathcal{E}(C(\alpha_{\Sigma}))$ where $\Sigma:\phi\to Y$ is a morphism in $\Rb$ defines a projective representation of $\Rb$ in \textbf{Hilb}.
                 
                 \begin{proof}  Let $\Sigma_1:Y_1\to Y_2$ and $\Sigma_2:Y_2\to Y_3$ be morphisms in $\Rb$. From corollary \ref{C2.4} we recall  \[\alpha_{\Sigma_2}\circ\alpha_{\Sigma_1}=\alpha_{\Sigma_2\circ\Sigma_1}.\] Now using proposition \ref{P4.1.6} to $\alpha_{\Sigma_2}$, $\alpha_{\Sigma_1}$ and $\alpha_{\Sigma_2\circ\Sigma_1}$ we get 
                                     \[\mathcal{E}(C(\alpha_{\Sigma_2}))\circ \mathcal{E}(C(\alpha_{\Sigma_1}))=c(\alpha_{\Sigma_2},\alpha_{\Sigma_1})\mathcal{E}(C(\alpha_{\Sigma_2\circ\Sigma_1}))\] where 
                                     \begin{multline}c(\alpha_{\Sigma_2},\alpha_{\Sigma_1})\\
                                                          =\dfrac{\det(\alpha_{\Sigma_1})^{\frac14}\cdot \det(\alpha_{\Sigma_1})^{\frac14}}{\det(\frac{(A+M)(\de_{Y_2}+m^2)^{-\frac12}}{2})^{\frac12}
                      \det(\mathcal{E}(C(\alpha_{\Sigma_2\circ\Sigma_1})))^{\frac14}}\dfrac{||\mathcal{E}(C(\alpha_{\Sigma_2})||\cdot||\mathcal{E}(C(\alpha_{\Sigma_1})||}{||\mathcal{E}(C(\alpha_{\Sigma_2\circ\Sigma_1}))||}
                                     \end{multline} which is nonzero. 
                \end{proof}
                                    
                \end{theorem}

                                   \hskip-4.5mmNow the idea is to  ``deprojectivize'' the projective representation constructed above to construct Free Scalar Theory.

                                     \hskip-4.5mmConsider $\Sigma:\varnothing\to Y$ and define \[E(\Sigma)=\frac{1}
             {\det_{\zeta}(\Delta_{\Sigma_1,D}+m^2)^{\frac12}\cdot \det_{\zeta}(2D_{\Sigma})^{\frac14}}\cdot  \frac{\mathcal{E}(C(\alpha_{\Sigma}))}{||\mathcal{E}(C(\alpha_{\Sigma}))||}.\] Here the notation $\de_{\Sigma, D}$ means the operator $\de_{\Sigma}$ with the Dirichlet boundary condition and $\det_{\zeta}$ is for the zeta regularized determinant (\cite{LF}). 
           
             \hskip-4.5mmMore generally, if \[\Sigma:Y_1\to Y_2\] is a morphism in $\Rb$, we think of \[\Sigma:\phi\to \overline{Y_1}\sqcup Y_2\] and then define $E(\Sigma)$ as before. This means $E(\Sigma)\in E(Y_1)^{\lor}\otimes E(Y_2)$ and consequently we get a Hilbert-Schmidt operator \[E(\Sigma):E(Y_1)\to E(Y_2).\] In fact, $E(\Sigma)$ will be a trace class operator as a consequence of \ref{T4.2} because we can always write $\Sigma$ as a composition of two morphisms in $\Rb.$ Also, by construction we have \[E(\sqcup_{i}\Sigma_i)=\otimes_i E(\Sigma_i).\]

                              \begin{lemma}\label{L3.3} Let $\Sigma: \phi\to Y$ then \[\det(\alpha_{\Sigma})=\dfrac{\det_{\zeta}(2D_{\Sigma})}{\det_{\zeta}(2(\Delta_{Y}+m^2)^{\frac12})}.\]
                              \begin{proof}  We note  \[2(\Delta_{Y}+m^2)^{\frac12}\alpha_{\Sigma}=2D_{\Sigma}.\] Recall that $\det(\alpha_{\Sigma})$ exists as $\alpha_{\Sigma}-I$ is trace class operator. Using  a well known fact ( If $A-I$ is trace class then $\det_{\zeta}(AB)=\det(A).\det_{\zeta}B$ proposition 2.20 \cite{Scott}), we get \[{\det}_{\zeta} (2(\Delta_{Y}+m^2)^{\frac12}\alpha_{\Sigma})={\det}_{\zeta}(2(\Delta_{Y}+m^2)^{\frac12})\det(\alpha_{\Sigma}).\] This gives the desired result $\det(\alpha_{\Sigma})=\dfrac{\det_{\zeta}(2D_{\Sigma})}{\det_{\zeta}(2(\Delta_{Y}+m^2)^{\frac12})}$.
                              \end{proof}
                              \end{lemma}

                              \hskip-4.5mmNow we prove the gluing law for the operators. 
                              
                               \begin{theorem}\label{T4.2} 
                                Let $\Sigma_1:Y_1\to Y_2$ and $\Sigma_2:Y_2\to Y_3$ be two morphisms in $\Rb$.

                               (i) There exists a non zero constant $C_{\Sigma_2,\Sigma_1}$ such that \[E(\Sigma_2)\circ E(\Sigma_1)=C_{\Sigma_2,\Sigma_1}E(\Sigma_2\circ\Sigma_1).\]

                               (ii) When $d$ is even, $C_{\Sigma_2,\Sigma_1}=1$.

                               \begin{proof}  
                               For $\Sigma_1:Y_1\to Y_2$, we write \[D_{\Sigma_1}= \left[\begin{smallmatrix} A & B\\ B^t &D\end{smallmatrix}\right] \hskip0.03in\text{where}\hskip0.03in D_{\Sigma_1}:C^{\infty}(Y_2)\oplus C^{\infty}(Y_1)\to C^{\infty}(Y_2)\oplus C^{\infty}(Y_1).\] Similarly for $\Sigma_2:Y_2\to Y_3$, we write \[D_{\Sigma_2}= \left[\begin{smallmatrix} K & L\\ L^t &M\end{smallmatrix}\right]\hskip0.03in\text{where}\hskip0.03in D_{\Sigma_2}:C^{\infty}(Y_3)\oplus C^{\infty}(Y_2)\to C^{\infty}(Y_3)\oplus C^{\infty}(Y_2).\]
                               
                                 \hskip-4.5mmWe note  
                                \begin{align*}&E(\Sigma_2)\circ E(\Sigma_1)\\
                                                     &=\dfrac{1}{det_{\zeta}(\Delta_{\Sigma_1,D}+m^2)^{\frac12}\det_{\zeta}(2D_{\Sigma_1})^{\frac14}}\\
                                                     &\cdot
                                                                        \dfrac{1}{det_{\zeta}(\Delta_{\Sigma_2,D}+m^2)^{\frac12}\det_{\zeta}(2D_{\Sigma_2})^{\frac14}}\\
                                                     &\cdot\frac{1}{\|\mathcal{E}(C(\alpha_{\Sigma_1}))\|}\cdot\frac{1}{\|\mathcal{E}(C(\alpha_{\Sigma_1}))\|}\mathcal{E}(C(\alpha_{\Sigma_2}))\circ\mathcal{E}(C(\alpha_{\Sigma_1}))                                                                      
                                \end{align*}

                                 \hskip-4.5mmBy proof of theorem \ref{Ch5:T4.1}, we have  
                                \begin{align*}&\frac{1}{\|\mathcal{E}(C(\alpha_{\Sigma_1}))\|}\cdot\frac{1}{\|\mathcal{E}(C(\alpha_{\Sigma_1}))\|} \E(C(\alpha_{\Sigma_2}))\circ\mathcal{E}(C(\alpha_{\Sigma_1}))\\
                                                    &= \dfrac{det(\alpha_{\Sigma_1})^{\frac14}det(\alpha_{\Sigma_2})^{\frac14}}{det(\alpha_{\Sigma_2\circ\Sigma_1})^{\frac14}det(\frac12(\Delta_{Y_2}+m^2)^{-\frac12}D_{\Sigma_1,\Sigma_2})^{\frac12}}
                                                    \cdot \frac{\mathcal{E}(C(\alpha_{\Sigma_2\circ\Sigma_1}))}{\|\mathcal{E}(C(\alpha_{\Sigma_2\circ\Sigma_1}))\|}                                                                 
                                \end{align*} where $D_{\Sigma_1,\Sigma_2}=A+M.$

                           \hskip-4.5mm Let us recall from the lemma \ref{L3.3} \begin{align*}\det(\alpha_{\Sigma_1})
                                                                                                            &=\frac{\det_{\zeta}(2D_{\Sigma_1})}{\det_{\zeta}(2(\Delta_{Y_1}+m^2)^{\frac12})\det_{\zeta}(2(\Delta_{Y_2}+m^2)^{\frac12})}
                                                                                        \end{align*}
                             
 \[\det(\alpha_{\Sigma_2})
                                                                                                            =\frac{\det_{\zeta}(2D_{\Sigma_2})}{\det_{\zeta}(2(\Delta_{Y_2}+m^2)^{\frac12})\cdot \det_{\zeta}(2(\Delta_{Y_3}+m^2)^{\frac12})},\] 
                                \[\det(\alpha_{\Sigma_2\circ\Sigma_1})
                                                                                                            =\frac{\det_{\zeta}(2D_{\Sigma_2\circ\Sigma_1})}{\det_{\zeta}(2(\Delta_{Y_1}+m^2)^{\frac12})\det_{\zeta}(2(\Delta_{Y_3}+m^2)^{\frac12})},\] and using argument of the same lemma             
                              
                             \[\det(\frac12(\Delta_{Y_2}+m^2)^{-\frac12}D_{\Sigma_1,\Sigma_2})=\frac{\det_{\zeta}(D_{\Sigma_1,\Sigma_2})}{\det_{\zeta}(2(\Delta_{Y_1}+m^2)^{\frac12})}.\]   
                             
                             \hskip-4.5mmUsing these relations, we see \[\frac{\det(\alpha_{\Sigma_1})^{\frac14}\cdot \det(\alpha_{\Sigma_2})^{\frac14}}{\det(\frac12(\Delta_{Y_2}+m^2)^{-\frac12}D_{\Sigma_1,\Sigma_2})^{\frac12}\cdot \det(\alpha_{\Sigma_2\circ\Sigma_1})^{\frac14}}=
\frac{\det_{\zeta}(2D_{\Sigma_1})^{\frac14}\cdot \det_{\zeta}(2D_{\Sigma_2})^{\frac14}}{\det_{\zeta}(D_{\Sigma_1,\Sigma_2})^{\frac12}\cdot \det_{\zeta}(2D_{\Sigma_3})^{\frac14}}.\]
                              This gives   
                              \begin{align*}&\frac{1}{\|\mathcal{E}(C(\alpha_{\Sigma_1}))\|}\cdot\frac{1}{\|\mathcal{E}(C(\alpha_{\Sigma_1}))\|}\E(C(\alpha_{\Sigma_2}))\circ\mathcal{E}(C(\alpha_{\Sigma_1}))\\
                              &=\frac{\det_{\zeta}(2D_{\Sigma_1})^{\frac14}\cdot \det_{\zeta}(2D_{\Sigma_2})^{\frac14}}{\det_{\zeta}(D_{\Sigma_1,\Sigma_2})^{\frac12}\cdot \det_{\zeta}(2D_{\Sigma_3})^{\frac14}}
                             \frac{\mathcal{E}(C(\alpha_{\Sigma_2\circ\Sigma_1}))}{\|\mathcal{E}(C(\alpha_{\Sigma_2\circ\Sigma_1}))\|} 
                              \end{align*}  Hence,                  
                              \begin{align*}E(\Sigma_2)\circ E(\Sigma_1)&=\frac{1}{\det_{\zeta}(\Delta_{\Sigma_1,D}+m^2)^{\frac12}\cdot \det_{\zeta}(\Delta_{\Sigma_2,D}+m^2)^{\frac12}
                                                                                                \cdot \det_{\zeta}(D_{\Sigma_1,\Sigma_2})^{\frac12}\cdot \det_{\zeta}(2D_{\Sigma_3})^{\frac14}}\\& \frac{\mathcal{E}(C(\alpha_{\Sigma_2\circ\Sigma_1}))}{\|\mathcal{E}(C(\alpha_{\Sigma_2\circ\Sigma_1}))\|}.
                                 \end{align*}
                                 Using the gluing formula for the zeta regularized determinants (\cite{B}, Theorem A or \cite{Lee2}, Corollary 1.3), we have \[{\det}_{\zeta}(\Delta_{\Sigma_2\circ\Sigma_1,D}+m^2)=K_{\Sigma_1,\Sigma_2}{\det}_{\zeta}(\Delta_{\Sigma_1,D}+m^2){\det}_{\zeta}(\Delta_{\Sigma_1,D}+m^2){\det}_{\zeta}(D_{\Sigma_1,\Sigma_2})\]  for some nonzero constant $K_{\Sigma_1,\Sigma_2}$. 
 Hence, \[E(\Sigma_2)\circ E(\Sigma_1)=\dfrac{1}{\sqrt{K_{\Sigma_1,\Sigma_2}}}E(\Sigma_2\circ\Sigma_1).\] Furthermore, when $d$ is even $K_{\Sigma_1,\Sigma_2}=1$ (\cite{Lee2}, Remark 1 after Corollary 1.3). 
                               \end{proof}
                                \end{theorem}
   \begin{corollary} Suppose that $\Sigma$ is a closed $d$-dimensional oriented Riemannian manifold and $d$ is even. Assume that $\Sigma=\Sigma_2\circ\Sigma_1$ in $\Rb$ where $\Sigma_1:\phi\to Y$ and $\Sigma_2:Y\to \phi.$ Then
   \[E(\Sigma)=\frac{1}{\det_{\zeta}(\Delta_{\Sigma}+m^2)^{\frac12}}.\]
   \end{corollary}  
    
    \hskip-4.5mm As suggested by the previous corollary, we complete the construction of the functorial free field theory by assigning to a closed oriented $d$ dimensional Riemannian manifold the number $E(\Sigma)=\frac{1}{\det_{\zeta}(\Delta_{\Sigma}+m^2)^{\frac12}}.$

     \subsection{More examples of Free Theory}\label{MFT}
                        
                        We describe an immediate generalization of previous examples which is to quantize a more general classical field theory. We take $\mathfrak{F}(\Sigma)$ to be sections of a real finite dimensional vector bundle $F$ on $\Sigma.$ Assume that $F$ comes with a metric $g^{F}$ and a compatible connection $\nabla^{F}.$ Let $d_{\nabla^{F}}:\Omega^{*}(M,F)\to \Omega^{*+1}(M, F)$ be the twisted deRham operator acting on $F$-valued forms. We are interested in $\Omega^{0}(M,F)$ where the action functional is defines as

            \[S(\phi)=\frac12\int_{\Sigma}\left(d_{\nabla^{F}}\phi\wedge\ast d_{\nabla^{F}}\phi\right)_{g^F}+m^2(\phi\wedge\ast\phi)_{g^F}\] where $\ast$ is the Hodge star operator.
            
           \hskip-4.5mm We denote the Hodge Laplacian acting on sections of $F$ by $\Delta_{\Sigma}$ i.e. $\Delta_{\Sigma}=d_{\nabla^{F}}^{*}\circ d_{\nabla^{E}}$ where $d_{\nabla^{F}}^{*}$ is the formal adjoint of $d_{\nabla^{F}}.$
            
            \hskip-4.5mmWe consider the semicategory $\mathcal{C}_{d}$ which is a straightforward generalization of $\Rb.$ An object in $\mathcal{C}_{d}$ is a quadruple $(Y, F, g^{F}, \nabla^{F})$. Here $Y$ is a $d-1$ dimensional closed oriented Riemannian manifold, $F$ is a real vector bundle over $Y$ of finite rank, $g^{F}$ a metric on $F$ and $\nabla^{F}$ is a flat metric connection. Let $(Y_i, F_i, g^{F_i}, \nabla^{F_i})$, $i=1,2$ be two objects of $\Cd.$ A morphism from $(Y_1, F_1, g^{F_1}, \nabla^{F_1})$ to $(Y_2, F_2, g^{F_2}, \nabla^{F_2})$ is a quadruple $(\Sigma, F, g^{F}, \nabla^{F})$ where $\Sigma: Y_1\to Y_2$ is a bordism in $\Rb$, F is a vector bundle on $\Sigma$ with $F|_{Y_i}=F_i$, $g^{F}$ is a metric on $F$ with $g^{F}|_{F_i}=g^{F_i}$ and $\nabla^{F}$ is a flat metric connection compatible with the connections $\nabla^{F_i}$ on $F_i.$ We assume the geometry of all the metrics and the connection $\nabla^{F}$ is of $product$ $type$ near $Y_i.$ We define the composition of morphisms to be the one given by gluing. 
                       
                \hskip-4.5mmLet $(Y, F, g^{F}, \nabla^{F})$ be an object of $\Cd.$ We define $W^{\frac12}(Y,F)$ to be completion of $\Cn(Y,F)$ with respect to the inner product \[(\phi,\psi)=\int_{Y}(\phi, (\Delta_{F}+m^2)^{\frac12}\psi)_{g^F}\, dvol(Y).\]

              \hskip-4.5mmLet $(\Sigma, F, g^{F}, \nabla^{F})$ be a morphism in $\Cd$ from $\phi$ to $(Y, F, g^{F}, \nabla^{F}).$ We define $D_{\Sigma}$, $\alpha_{\Sigma}$ and $\E(C(\alpha_{\Sigma}))$ as before (see section 2 and section 3).  

              \hskip-4.5mmTo an object $(Y, F, g^{F}, \nabla^{F})$ in $\Cd$, assign the Hilbert space \[E((Y, F, g^{F}, \nabla^{F}))=\Sm W^{\frac12}(Y,F)^{\lor}\] and to a morphism $(\Sigma, F, g^{F}, \nabla^{F})$ from $\phi$ to $(Y, F, g^{F}, \nabla^{E})$ assign the vector 
                                                \[\frac{1}
             {\det_{\zeta}(\Delta_{\Sigma_1,D}+m^2)^{\frac12}\cdot \det_{\zeta}(2D_{\Sigma})^{\frac14}}\cdot  \frac{\mathcal{E}(C(\alpha_{\Sigma}))}{||\mathcal{E}(C(\alpha_{\Sigma}))||}\in E((Y, F, g^{F}, \nabla^{F})) .\] This assignment defines a projective Functorial QFT for all $d$ and Functorial QFT when $d$ is even as above. 

           \hskip-4.5mm In particular, if we take the trivial vector bundle $\underline{\R}^N$ then we get a Functorial QFT with target $\R^{N}.$

\section{Appendix}
\subsection{Bosonic Fock space}\label{4.1}

                  \begin{note}
                
                  In what follows we assume the Hilbert spaces in consideration are separable.
                
                \end{note}
        
             \begin{definition} 
             
                     The $Bosonic$ $Fock$ space of a Hilbert space $H$ is the Hilbert space direct sum
                    \[\bigoplus_{n=0}^{\infty}\sn(H)=\{(\alpha_n)_{n=0}^{\infty}:\alpha_n\in \sn H \hspace{0.05in}\text{with}\hspace{0.05in}\sum_{n=0}^{\infty}\|\alpha_n\|^2<\infty\}\] where $\sn(H)$ is the closed subspace of $H^{\otimes n}$ that is invariant under the action of permutation group $S_n.$ We note $\sn(H)$ is the closure of the subspace spanned by vectors of the form \[h_1\otimes_{s}\dots\otimes_{s}h_n=\frac{1}{\sqrt{n!}}\sum_{\sigma\in S_n}h_{\sigma(1)}\otimes\dots\otimes h_{\sigma(n)}.\]
                    We use $\Sm(H)$ to denote the Bosonic Fock space of $H$. We also use the notation $h_1\cdot h_2\dots h_n$ for $h_1\otimes_{s}\dots\otimes_{s}h_n.$
                   
                     \end{definition} 
                     
                       \hskip-4.5mmLet $H$ be a real Hilbert space and let $A:H\to H$ be a symmetric Hilbert Schmidt linear operator. Define a map $S(A):H\times H\to \R$ by \[S(A)(u\otimes v)= \hskip0.02in<Au,v>.\] Since $A$ is symmetric and Hilbert Schmidt, we see $S(A)\in\smt(H^{\lor}).$ Moreover, if $O:H\to H$ is an orthogonal operator then $S(OAO^{-1})=S(A).$ By abuse of notation we write $A$ for $S(A)$.
                     
                     \begin{definition}
                     
                       Let $H$ be a finite dimensional real Hilbert space and  $A:H\to H$ be a symmetric linear operator. We define
                     \[\E(A)=\exp(\frac12 S(A)) \] where \[\exp(\alpha)=\sum_{n=0}^{\infty}\frac{\alpha^n}{n!}\]
                     for all $\alpha\in \sk H^{\lor}.$
                     
                     \end{definition}
                     
                     \begin{prop}\label{P4.1} Assume that $H$ is finite dimensional and $A:H\to H$ a symmetric linear operator. Then $\E(A)\in \Sm(H^{\lor})$  if and only if $\|A\|<1.$ If $\|A\|<1$, then \[\|\E(A)\|^{2}=\frac{1}{\sqrt{det({I-A^2})}}.\]
                     
                     \end{prop}

                      \hskip-4.5mmProposition \ref{P4.1} is a direct consequence of the following elementary lemma.
                     
                     \begin{lemma} 
                     
                        Assume that $H$ is one dimensional and $x\in H^{\lor}$ such that $\|x\|=1$. Let $a\in \R$. Then $\exp(\frac12ax^2)\in \Sm(H^{\lor})$  if and only if $|a|<1.$ If $|a|<1$ then $\|\exp(\frac12ax^2)\|^{2}=\frac{1}{\sqrt{{1-a^2}}}.$
                     
                     \begin{proof}
                     
                       We note $\|x^n\|^2=n!$. Hence,
                     \[\|\exp(\frac12 ax^2)\|^2=\sum_{n=0}^{\infty}\frac{a^{2n}(2n)!}{2^{2n}(n!)^2}.\] We know that the series $\sum_{n=0}^{\infty}\frac{a^{2n}(2n)!}{2^{2n}(n!)^2}$ converges if and only if $|a|<1$ and when it converges the sum is $\frac{1}{\sqrt{{1-a^2}}}.$
                     
                     \end{proof}
                     \end{lemma}

                     \hskip-4.5mmAssume that $H$ is finite dimension real Hilbert space and let $A$ and $B$ be symmetric linear operators on $H$ of norm smaller than one.  A direct computation shows 
                \begin{eqnarray*}<\E(A),\mathcal{E}(B)>=\det(1-AB)^{-\frac12}
                \end{eqnarray*} and  
               \begin{eqnarray*}
               \begin{array}{lll}&\|\E(A)-\E(B)\|^2
               =\det(1-A^2)^{-\frac12}+\det(1-B^2)^{-\frac12}-2(\det(1-AB)^{-\frac12}).
               \end{array}
               \end{eqnarray*}
                        
           \hskip-4.5mm As a immediate consequence of this calculation we get the assignment $A\mapsto \E(A)$ is continuous on the space of symmetric operators in $H$ which have norm smaller than one.  

           \hskip-4.5mmNow we move on to the infinite dimensional case. Let $\mathcal{Z}(H)$ denote the space of symmetric Hilbert-Schmidt linear operators on $A:H\to H$ such that the operator norm of $A$ is smaller than one. We note $\mathcal{Z}(H)$ is an open subset of the space of Hilbert-Schmidt operators on $H.$ Now using the fact each Hilbert-Schmidt operator can be approximated by finite rank operators in the space of Hilbert Schmidt operators we have a continuous map $\E:\mathcal{Z}(H)\to \Sm H^{\lor}$ given by 
\[\E(A)=\exp(\frac12 S(A)).\] It also follows \[<\E(A),\mathcal{E}(B)>
                                                                      =\det(1-AB)^{-\frac12}\] for all $A,B\in\mathcal{Z}(H).$

\hskip-4.5mmConsider the map $C:(0,\infty)\to (-1,1)$ given by $x\mapsto (1-x)(1+x)^{-1}.$ It is easy to see $C$ is a homeomorphism. Now we use $C$ to define the $Cayley$ $transform$ of an operator. 
                      
                      \begin{definition}\label{def10}
                       Let $H$ be a real Hilbert space and let $A:H\to H$ be a continuous positive operator. The $Cayley$ $transform$ of $A$ is defined by  $C(A)=(I-A)(I+A)^{-1}$. We note $C(A)\in\mathcal{Z}(H)$ whenever $A$ is a positive operator on $H$ where $\mathcal{Z}(H)$ is the set of symmetric operator on $H$ with norm smaller than 1.
                     
                       \end{definition}

                        \hskip-4.5mm Let $H$ be a real Hilbert space. The set $\S(H)$ denotes the set of continuous positive linear operators on $H$ such that $A-I$ is Hilbert-Schmidt. We note each $A\in\S(H)$ has a bounded inverse.

             \hskip-4.5mmThere is an obvious map $\S(H)$ to the Hilbert space of Hilbert Schmidt operators on $H$ given by $A\mapsto A-I$. We give $\S(H)$ the weakest topology such that the map $A\mapsto A-I$ is a continuous map. In other words, $A_n$ converges to $A$ in $\S(H)$ if and only if $A_n-I$ converges to $A-I$ with respect to the Hilbert Schmidt norm. The proof of the following lemma is obvious.
                      \begin{lemma}
                      
                        $C:\S(H)\to \mathcal{Z}(H)$ is a homeomorphism. 
                      
                      \end{lemma}

                     \hskip-4.5mmAssume that $H_1$, $H_2$ and $H_2$ are finite dimensional real Hilbert spaces. Let $\CA_1\in\S(H_2\oplus H_1)$ and $\CA_2\in \S(H_3\oplus H_2).$ We write  \[\CA_1=\left[\begin{smallmatrix} A & B\\ B^t &D\end{smallmatrix}\right] \hskip0.05in\text{and}\hskip0.05in\CA_2=\ \left[\begin{smallmatrix} K & L\\ L^t &M\end{smallmatrix}\right]\] and define 

                  \begin{equation}
\CA_2\circ\CA_1=\left[\begin{smallmatrix} K-L(A+M)^{-1}L^t & -L(A+M)^{-1}B\\ -B^t(A+M)^{-1}L^t &D-B^t(A+M)^{-1}B\end{smallmatrix}\right]
                 \end{equation}

                       \hskip-4.5mmIn fact $\CA_2\circ\CA_1\in \S(H_3\oplus H_1)$ and the operation $\circ$ is associative and continuous (Lemma 3.2.11 in \cite{SK}).  

                       \hskip-4.5mm Now using the Cayley transform we can define $\circ:\mathcal{Z}(H_3\oplus H_2)\times\mathcal{Z}(H_2\oplus H_1)\to \mathcal{Z}(H_3\oplus H_1)$ that makes the following diagram commutative:

                  \begin{center}
                         \begin{tikzpicture}
                         
                         \matrix (m) [matrix of math nodes, row sep=2em,
                          column sep=1em, text height=1.5ex, text depth=0.25ex]
{\S(H_3\oplus H_2)\times \S(H_2\oplus H_1)&& \S(H_3\oplus H_1) \\
   \mathcal{Z}(H_3\oplus H_2)\times\mathcal{Z}(H_2\oplus H_1)&& \mathcal{Z}(H_3\oplus H_1)&&\\ };

\path[->]
(m-1-1) edge node[above]{$\circ $}(m-1-3);

\path[->]
 (m-1-1)edge (m-2-1)
(m-1-3) edge (m-2-3);
\path[->]
(m-2-1) edge node[below]{$\circ $}(m-2-3);
                         \end{tikzpicture}
                         \end{center}

\hskip-4.5mmLet $\CA_1\in \S(H_2\oplus H_1)$ then $\E(C(\CA_1))\in\Sm (H_2^{\lor}\oplus H_1^{\lor}).$  Note that  $\E(C(\CA_1))$ can be canonically identified with a Hilbert-Schmidt operator \[\Sm H_1^{\lor}\to \Sm H_1^{\lor}.\]  We use $\E(C(\CA_1))$ to denote this Hilbert-Schmidt operator by abusing the notation. Similarly $\CA_2\in\S(H_3\oplus H_2)$ induces a Hilbert-Schmidt operator \[\E(C(\CA_2)):\Sm H_2^{\lor}\to \Sm H_3^{\lor}.\] Now, we are ready for a key  proposition which was used in proposition 3.2.  

         \begin{prop}\label{P4.4}
         Let $\CA_1$ and $\CA_2$ as above. Then $\mathcal{E}(C(\mathcal{A}_2))\circ \mathcal{E}(C(\mathcal{A}_1))=c(\mathcal{A}_2,\mathcal{A}_1)\mathcal{E}(C(\mathcal{A}_2\circ\mathcal{A}_1))$ where  
                      \begin{align*} &c(\mathcal{A}_2,\mathcal{A}_1)=\dfrac{\det(\mathcal{A}_1)^{\frac14}\cdot \det(\mathcal{A}_2)^{\frac14}}{\det(\frac{A+M}{2})^{\frac12}
                      \det(\mathcal{A}_2\circ\mathcal{A}_1)^{\frac14}}\dfrac{||\mathcal{E}(C(\mathcal{A}_1))||\cdot||\mathcal{E}(C(\mathcal{A}_2))||}{||\mathcal{E}(C(\mathcal{A}_2\circ\mathcal{A}_1))||}
                     \end{align*}
                      and the norms are taken in the Fock spaces.
                     
                     \begin{proof}  
                     Using proposition \ref{P4.10} we observe \[\frac{\E(C(\CA_i))}{\|\E(C(\CA_i))\|}= \det(\CA_i)^{\frac14}Q\circ L_{\CA_i}\circ Q^{-1}\]  and 
                     
                     \[L_{\CA_2}\circ L_{\CA_1}=C'(\CA_2,\CA_1) L_{\CA_2\circ \CA_1}\] where $C'(\CA_2,\CA_1)=\det(\frac12(A+M))^{-\frac12}.$ Now,
                     \begin{align*} \E(\CA_2)\circ \E(\CA_1)&=\det(\CA_1)^{\frac14}.\det(\CA_1)^{\frac14}.\\
             & \hskip0.2in\|\E(C(\CA_2))\|.\|\E(C(\CA_1))\| \\
             &  \hskip0.2in Q\circ (L_{\CA_2}\circ L_{\CA_1})\circ Q^{-1}\\
             &=\frac{\det(\CA_1)^{\frac14}.\det(\CA_1)^{\frac14}}{\det((\frac12A+M))^{\frac12}}.\\
             &\hskip0.2in\|\E(C(\CA_2))\|.\|\E(C(\CA_1))\|
             \hskip0.01in Q\circ (L_{\CA_2\circ\CA_1})\circ Q^{-1}\\
             &=C(\CA_2, \CA_1)\E(C(\CA_2\circ\CA_1)).
                     \end{align*}
                     \end{proof}
                     \end{prop}

\subsection{Segal-Ito isomorphism finite dimensional case}

Let $\mu$ be a standard Gaussian measure on $\R$. We recall it defines an inner product on $\R^{\lor}$. There is a unique isomorphism $Q:\Sm(R^{\lor})\to L^2(\R,\, d\mu)$ which on $\sn(R^{\lor})$ is given by \[Q(x^n)=h_n\] where $x\in \R^{\lor}$ is dual to $1\in\R$ and $h_n$ is the degree $n$ Hermite polynomial.

   \hskip-4.5mm We recall the following which we will use later.                      
                         \begin{fact} $e^{ax-\frac12a^2}=\sum_{n=0}^{\infty}\frac{1}{n!}a^{n}h_n.$
                         \begin{proof} We note  
                                \begin{align*}
                         e^{ax-\frac12a^2}&=e^{\frac12x^2-\frac12(x-a)^2}\\
                         &=e^{\frac12x^2}e^{-\frac12(x-a)^2}.
                               \end{align*}
                               Now we write \[e^{-\frac12(x-a)^2}=\sum_{n=0}^{\infty}\frac{a^n}{n!}\frac{d^n}{da^n}e^{-\frac12(x-a)^2}|_{a=0}\] 
                           which is the Maclaurin series expansion in the variable $a$. We note \[\frac{d^n}{da^n}e^{-\frac12(x-a)^2}|_{a=0}=(-1)^n\frac{d^n}{dx^n}e^{-\frac12x^2}.\] This gives us  \[e^{ax-\frac12a^2}=\sum_{n=0}^{\infty}\frac{a^n}{n!}(-1)^ne^{\frac12x^2}\frac{d^n}{dx^n}e^{-\frac12x^2}=\sum_{n=0}^{\infty}\frac{1}{n!}a^{n}h_n\]
                         \end{proof}
                          \end{fact}
                   \hskip-4.5mm Hence, \[Q(\exp(ax))=e^{ax-\frac12a^2}\] where $e^{ax-\frac12a^2}$ is considered as function on $\R.$ It also follows \[Q(\exp h)=e^{h-\frac12||h||^2}\] for all $h\in \R^{\lor}$implying $Q$ is unique. 
                    
                    \hskip-4.5mmWe denote the complex valued $L^2$ functions by $L^2(\R,\mu,\C)$. Then $Q$ induces an unitary map $Q:\Sm(\C^{\lor})\to L^2(\R,\mu,\C).$ 
                       
                          \hskip-4.5mm Let $H$ be a finite dimensional real vector space and $\mu_{H}$ a centered  Gaussian measure on $H$. The Gaussian measure induces an inner product on $H$. Then the  discussion above for the one dimensional case can be generalized in a routine way to construct the unique isomorphism  
             \[Q:\Sm(H^{\lor})\to L^2(H,\, \mu_{H})\] 
with the property 
            \[Q(\exp h)=e^{h-\frac12\|h\|^2}.\] 
As in the one dimensional case, $Q$ induces an unitary map 
            \[Q:\Sm(H_{\C}^{\lor})\to L^2(H,\, \mu_{H},\C)\] 
where $H_{\C}$ is the complexification of $H$ and $L^2(H,\, \mu_{H},\C)$ is the $L^2$ space of complex valued functions.
                          This map $Q$ can be thought as a ``backward heat'' operator which maps polynomials to $L^2$ functions. Next, we attempt to justify this assertion.

                      \hskip-4.5mmLet us identify $H$ with $\R^n$ such that $\mu_{H}$ corresponds to the standard Gaussian measure $\mu_n$ on $\R^n$. Let $\nu_{n}$ denote the measure on $\C^n$ with the density $\pi^{-n}e^{-\|z\|^2}$ with respect to the standard Lebesgue measure on $\C^n=\R^{2n}$. We use $\mathcal{H}(\C^n)$ to denote the space of holomorphic functions on $\C^n$ and $\mathcal{H}L^2(\C^n,\nu_{n})$ to denote the space of $L^2$-holomorphic functions on $\C^n$.

                       \hskip-4.5mmThere is a well known unitary map \cite[Theorem 2.3]{GM} $S:L^2(\R^n,\,d\mu,\C)\to \mathcal{H}L^2(\C^n,\mu_{\C})$ defined by \[Sf(z)=e^{-\frac12 <z, z>} \int_{\R^n}f(u)e^{<z, u>}\,d\mu_n (u)=\int_{\R^n}B_n(z,u)f(u)\,d\mu_{n}(u)\] where \[B_n(z,u)=e^{-\frac12<z,z>+<z,u>}.\] 
                       Here $<,>$ is the complex bilinear extension of the standard inner product on $\R^n.$ The map $S$ takes Hermite polynomials to the $z^{\alpha}$s. Note that
                           \begin{equation} \label{eq3} Sf(z)=(2\pi)^{-\frac{n}{2}}\int_{\R^n}f(z-u)e^{-\frac12<u,u>}\,du.
                           \end{equation}

                         \hskip-4.5mm Identifying $\mathcal{H}L^2(\C^n,\mu_{\C})$ with $\Sm((\C^n)^{\lor})$ using the tautological unitary map $T$ called the $Taylor$ map 
                        \[T:\mathcal{H}L^2(\C^n,\mu_{\C})\to \Sm((\C^n)^{\lor})\] and identifying $S$ with $T\circ S$, it can be easily checked that $Q$ and $S$ are inverse of each other.
                        In fact, $Q$ from $\mathcal{H}L^2(\C^n,\nu_n)$ to $L^2(\R^n,\mu_n,\C)$ is given by
                         \[QF(x)=\int_{\C^n}B(\bar{z},x)F(z)\,d\nu_n(z).\]

 \hskip-4.5mm We also observe that the following diagram is commutative.                 
                      \begin{center}
                         \begin{tikzpicture}
                         
                         \matrix (m) [matrix of math nodes, row sep=1.5em,
                         column sep=1em, text height=1.5ex, text depth=0.25ex]
{L^2(\R^n,\,d\mu)&&\Sm({(\R^n)}^{\lor})\\
    L^2(\R^n,\,d\mu, \C)&&\Sm({(\C^n)}^{\lor})& &\\ };

\path[->]
(m-1-1) edge node[above]{$S$}(m-1-3);

\path[->]
 (m-1-1)edge (m-2-1)
(m-1-3) edge (m-2-3);
\path[->]
(m-2-1) edge node[below]{$S$}(m-2-3);
                         \end{tikzpicture}
                         \end{center}

\hskip-4.5mmLet $k(y,x)\in L^2(\R^m\times \R^n, \mu_m\otimes\mu_n).$ Then $k$ induces a Hilbert Schmidt operator \[L_{k}: L^2(\R^n, \mu_n)\to L^2(\R^m,\mu_{m})\] given by 
                            \[L_kf(y)=\int_{\R^n}k(y,x)f(x)\, d\mu_{n}(x).\] Furthermore $L_{k}$ induces a Hilbert Schmidt operator \[Q^{-1}\circ L_{k}\circ Q:\mathcal{H} L^2(\C^n, \nu_{n})\to \mathcal{H}L^2(\C^m, \nu_{m})\] where $Q:\mathcal{H}L^2(\C^n, \nu_{n})\to L^2(\R^n,\mu_n)$ and $Q^{-1}:L^2(\R^m,\mu_m)\to \mathcal{H}L^2(\C^m, \nu_{m}).$ We are interested in computing the kernel of the operator $Q^{-1}\circ L_{k}\circ Q$ explicitly. Let $K(w,z)$ denote the image of $k(y,x)$ under the isomorphism $Q^{-1}:L^2(\R^m\times R^n, \mu_m\otimes \mu_n)\to \mathcal{H}L^2(\C^{m+n},\nu_{m+n}).$
                           
                            \begin{prop}\label{P4.10}$Q^{-1}\circ L_{k}\circ QF(w)=\int_{\C^n} K(w,\overline{z})F(z)\nu_n(z).$ 
                            \begin{proof} We note  
                            \begin{equation}\label{EQ9}
                             Q^{-1}\circ L_{k}\circ QF(w)=\int_{\R^m}\int_{\R^n}\int_{\C^n}B_n(\bar{z},x)B_m(w,y)k(y,x)F(z)\,d\nu_n(z)\,d\mu_n(x)\, \,d\mu_m(y).\end{equation} 
                              Interchanging the order of integration (See for example Proposition 1.81 \cite{F}) and using the fact   \[B_{m+n}(w,z),(y,x)=B_m(w,y)B_n(z,x),\] we observe  
                            \[Q^{-1}\circ L_{k}\circ QF(w)=\int_{\C^n}\int_{\R^m}\int_{\R^n}B_{m+n}(w,\bar{z}),(y,x)k(y,x)\,d\mu_n(x)\,d\mu_m(y)F(z)d\nu_n(z)\]  when $F$ is a polynomial. Now we can conclude 
                            \[Q^{-1}\circ L_{k}\circ QF(w)=\int_{\C^n}K(w,\bar{z})F(z)\, d\nu_n(z).\] Since all the maps in the consideration are continuous it is sufficient to check when $F$ is a polynomial.

                            \end{proof}
                            \end{prop}

Max Planck Institute for Mathematics,
Bonn, Germany\\
$Email:$ \text{skandel1@alumni.nd.edu}

\end{document}